\documentclass[cameraready]{Interspeech}

\title{UrduSpeech: A 156-Hour Urdu Speech Corpus with 12-Dimension Paralinguistic Annotations}

\author[affiliation={1}, orcid=0000-0003-0376-5247]{Attia Nafees ul}{Haq}
\author[affiliation={1}, orcid=0009-0003-6722-2417]{Zeyu}{Zhu}
\author[affiliation={1}, orcid=0009-0003-8727-2411]{Jingbin}{Hu}
\author[affiliation={1}, orcid=0009-0005-2049-9501]{ChunJiang}{He}
\author[affiliation={1}, correspondingauthor]{Lei}{Xie}

\address{
    $^1$ Audio, Speech and Language Processing Group (ASLP@NPU),Northwestern Polytechnical University 
}

\email{attianafees@mail.nwpu.edu.cn, lxie@nwpu.edu.cn}

\keywords{under-resourced languages, paralinguistics, code-switching, automatic speech recognition, urdu, corpus curation}

\usepackage{comment}

\begin{document}

\maketitle

\begin{abstract}
Despite 230 million speakers, Urdu remains critically under-resourced in speech technology.We introduce UrduSpeech: large high-fidelity Urdu corpus comprising 156 hours of audio with 12-dimension paralinguistic metadata, encompassing US-Std,US-CS,US-EngPk.To address Right-To-Left script constraints and frequent code-switching, we developed UrduSpeech a LLM-driven pipeline to curate data across 12 diverse categories, including news, drama and rare literary forms like \textit{Bait-Bazi}. We also release a 9-hour US-benchmark set, manually corrected by native annotators to serve as a standard. Human quality assessment of the primary 156 hours corpus yielded a Mean Opinion Score (MOS) of 4.6 ($\sigma = 0.7$) with inter-rater reliability confirmed by a 0.68 Cohen’s Kappa, validating our curation pipeline's 97.6\% confidence score. The corpus maintains a 60/40 gender balance across 71,792 utterances.Our work represents a significant leap toward linguistic inclusivity in global AI. The corpus and code are open-sourced, and  demo page is available. \footnote{ \url{https://interspeech-urdu-demo.github.io/Urdu-corpus-demo/}}
\end{abstract}

\section{Introduction}
In recent years, the digital preservation of languages within the AI landscape has become a cornerstone of linguistic equality \cite{blasi2022systematic}. Yet, despite Urdu’s global significance and its vast diaspora, it remains remarkably under-resourced in the context of multimodal foundation models and Speech LLMs. Recent benchmarks highlight a persistent performance gap in Urdu ASR \cite{arif2025we}, primarily due to the lack of specialized tools capable of navigating Urdu’s unique challenges: its Right-to-Left (RTL) Perso-Arabic script \cite{bandarupalli2025towards}, the ubiquity of Urdu-English code-switching \cite{sharif2024survey, sadeqi2023cs}, and its acoustic proximity to Hindi \cite{daud2017urdu}. While large-scale initiatives like Omnilingual ASR \cite{omnilingual2025omnilingual} and Common Voice \cite{ardila2020common} have expanded coverage, specialized resources for nuanced tasks like Machine Reading Comprehension \cite{kazi2026uquad+}, Deepfake detection \cite{owais2026deepfake}, and Speech Emotion Recognition \cite{dar2026cross} remain scarce.

Motivated by the effectiveness of the WenetSpeech-Yue \cite{li2025wenetspeech} and WenetSpeech-Chuan \cite{dai2025wenetspeech} pipelines, we developed a specialized solution for the Urdu-English paradigm. We build upon foundational datasets including ARL Urdu \cite{LDC_ARL_Urdu_2007}, CLE Pakistan \cite{cle_urdu_speech_2016}, and LDC-IL \cite{LDCIL_Urdu_2023} while addressing the critical shortage of high-fidelity data in modern Urdu TTS \cite{khan2024overcoming} and ASR. Our primary motivation is the preservation of Standard Pakistani Urdu and its specific acoustic nuances, particularly the phonetic identity of Pakistani-accented English \cite{sarfraz2010phonological}. We introduce UrduSpeech, a 156-hour corpus designed to bridge the digital divide through accurate linguistic representation.

UrduSpeech\footnote{Ethical Statement: All data sourced from public repositories; no personal identifiers retained. Content is non-political/religious and adheres to local cultural norms.} bridges the "in-the-wild" gap \cite{nagrani2017voxceleb} through 12-layer paralinguistic metadata across 12 categories, including rare literary forms like \textit{Bait-Bazi}. This granular labeling of accent, emotion, and vocal texture allows for high-resolution error analysis across 71,792 utterances while maintaining a balanced 60/40 gender distribution. Such a framework, inspired by standard computational paralinguistic challenges \cite{schuller2013computational}, coupled with a 9-hour manually-verified benchmark for the US-Std, US-CS, and US-EngPK subsets, establishes a rigorous new ground truth for future speech processing research in under-resourced Perso-Arabic languages. The key contributions of our research can be summarized as follows:

\begin{itemize}
\item \textbf{UrduSpeech Pipeline:} A robust framework designed to filter raw audio, perform speaker diarization, and handle RTL script constraints while differentiating between Hindi and Urdu in code-switched environments.
\item \textbf{Benchmarking SOTA Speech LLMs:} An in-depth evaluation of Gemini 2.5 Pro \cite{comanici2025gemini}, Whisper-large-v3 \cite{radford2023robust}, and OmniASR-LLM-1 \cite{omnilingual2025omnilingual} to establish a baseline for high-fidelity transcription and paralinguistic annotation.
\item \textbf{US-Benchmark Set:} A 9-hour benchmark comprising US-Std, US-CS, and US-EngPK audios across 12 categories, manually validated by native annotators with 12-dimension paralinguistic metadata.
\item \textbf{UrduSpeech Corpus:} A 156-hour corpus consisting of 59.2h of US-Std, 89.4h of US-CS, and 7.3h of US-EngPk across 71,792 utterances. It includes comprehensive paralinguistic labels (emotion, texture, accent) verified by native speakers.
\end{itemize}

\begin{figure*}[tp]
  \centering
  \includegraphics[width=\linewidth, trim={0.5cm 7.5cm 0.5cm 7cm}, clip]{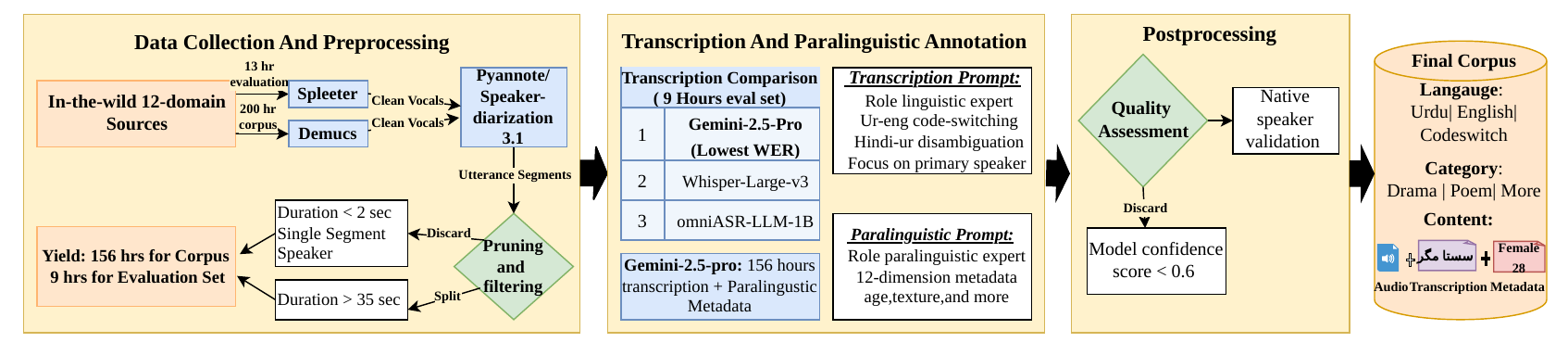}
  \caption{Overview of the UrduSpeech data curation pipeline.}
  \label{fig:pipeline}
\end{figure*}
\section{Model selection and benchmark set}

Prior to large-scale development, we conducted a 13-hour audio pilot study across 12 categories, including poetry, news, and vlogs. We gathered this raw audio "in-the-wild" and processed it according to our curation pipeline stage 1 as show in the figure \ref{fig:pipeline}. We utilized Spleeter \cite{hennequin2020spleeter} for noise removal and Pyannote \cite{bredin2020pyannote} for speaker diarization. To ensure high data quality, we discarded single-speaker clips and segments shorter than two seconds. Additionally, all audio clips were capped at a maximum duration of 35 seconds to optimize downstream transcription performance. This preprocessing resulted in our 9-hour, manually-verified US-benchmark set.

\subsection{Transcription model selection}
We compared three models for transcription: Whisper-v3 
 \cite{radford2023robust}, as it is the most commonly used model for Urdu; the recently released OmniASR-LLM-1B  \cite{omnilingual2025omnilingual}, which supports 1,600 languages and classifies Arab-Urdu as a high-resource language; and Gemini-2.5-Pro  \cite{comanici2025gemini} for its prompt engineering abilities and semantic awareness. We normalized and evaluated the outputs using JiWER \cite{morris2004and} against our native annotator ground truth; the results are displayed in Table \ref{tab:results}.

\begin{table}[htbp] 
  \caption{Word Error Rate (WER) performance across segments with  and without code-switching(CS)  for evaluated transcription models}
  \label{tab:results}
  \centering
  \small
  \setlength{\tabcolsep}{2pt} 
  \begin{tabular}{ l c c c c c }
    \toprule
    \textbf{Model}& \multicolumn{2}{c}{\textbf{Without CS}} & \multicolumn{2}{c}{\textbf{With CS}} & \textbf{Semantic} \\
    & \multicolumn{2}{c}{\textbf{(avg)}} & \multicolumn{2}{c}{\textbf{(avg)}} & \textbf{Awareness} \\
    \cmidrule(lr){2-3} \cmidrule(lr){4-5}
    & \textbf{WER} & \textbf{CER} & \textbf{WER} & \textbf{CER} & \\
    \midrule
    Whisper-Large-v3     & 0.289 & 0.185 & 0.532 & 0.556 & No \\
    0mniASR-LLM-1B     & 0.295 & 0.180 & 0.499 & 0.416 & No \\
    Gemini-2.5-Pro  & \textbf{0.023} & \textbf{0.017} & \textbf{0.028} & \textbf{0.018} & \textbf{Yes} \\
    \bottomrule
  \end{tabular}
\end{table}

As seen in the table, the average WER difference between Whisper-large-v3, OmniASR-LLM-1b, and Gemini-2.5-pro is quite significant. Upon further investigation, we deduced the following reasons: 
\begin{itemize}
    \item \textbf{OmniASR-LLM-1B:} Produced \textbf{hallucinations} in Arabic or Persian and exhibited \textbf{word-looping} on code-switched or accented segments.
    \item \textbf{Whisper-large-v3:} Failed on code-switched audio by \textbf{transliterating} or \textbf{translating} English into Urdu script rather than maintaining literal content.
    \item \textbf{Gemini-2.5-Pro:} Outperformed the others due to its \textbf{semantic awareness} and \textbf{targeted prompting}, which ensured Arab-Urdu script fidelity and annotated 12 paralinguistic labels such as age, texture, tone, and accent.
\end{itemize}
  
\subsection{US-Benchmark evaluation set and annotation}
We established the \textbf{3.4GB} 9-hour US-benchmark set to serve as our standard for error analysis. To make sure the ground truth was as accurate as possible, our native annotators went through and manually corrected all the Gemini model-generated transcriptions. This allowed us to fix subtle errors in code-switching and manually correct instances where the model output was in Hindi script instead of Urdu. In addition to the transcription, we used \textbf{Gemini 2.5 Pro} to tag each audio segment with 12 paralinguistic labels, such as pitch, rhythm, emotion, and accent. his metadata framework enables high-resolution analysis of how ASR performance fluctuates across diverse vocal characteristics and the Us-Std, US-CS, and US-EngPk subsets.

\section{UrduSpeech corpus curation pipeline}
Building on our US-benchmark set pilot, we scaled the corpus development into a multi-stage pipeline, as illustrated in Figure~\ref{fig:pipeline}. We further incorporated audio format metadata (short vs. long form) and integrated model confidence scores alongside quality assessments conducted by native annotators.

\subsection{Data collection and preprocessing}
We gathered \textbf{200 hours} of in-the-wild audio from YouTube and archival Pakistan Television (PTV) logs spanning the 1980s to the present, ensuring acoustic diversity across four decades. This collection spans media-trained and non-professional speakers, including vlogs, street interviews, and overseas Pakistanis, to capture authentic regional dialects and accent shifts in code-switched environments.

For audio preprocessing, we transitioned to the Demucs model \cite{defossez2019demucs} for more efficient source separation and utilized Pyannote 3.1 for speaker diarization. To maintain global speaker ID consistency, we adopted a \textit{one-file-at-a-time} approach followed by manual global alignment. Finally, we applied a strict pruning protocol: removing segments under 2 seconds or those from single-segment speakers and splitting clips exceeding 35 seconds. This resulted in UrduSpeech, containing 71,792 diarized clips across Us-Std, US-CS, and US-EngPk subsets, discarding 44 hours of residual noise.

\subsection{Gemini prompt engineering and data segmentation}
To handle the transcription and paralinguistic labeling, we developed a two-stage strategy using Gemini 2.5 Pro. First, we designed a transcription prompt that acted as an expert transcript specialist, strictly forbidding Hindi/Devanagari script to prevent script mixing. For code-switching, we forced a {\textit{literal transcription} constraint so the model would switch scripts mid-sentence to match the acoustic transition rather than translating. \

The second stage involved a paralinguistic analysis prompt covering 12 attributes like pitch, texture, and rhythm. We purposefully forbade the use of generic words like moderate or neutral to force the model to identify specific nuances, such as husky texture. We also instructed it to focus on the primary speaker despite the South Asian environmental noise.

To ensure corpus integrity, we implemented a rigorous filtering protocol based on these confidence scores, discarding any segment below $0.6$. Approximately 98\% of the data (71,101 segments) fell into the \textbf{Highly Accurate} category ($>0.9$), while only a small fraction fell into the \textbf{Reliable}, \textbf{Good}, or \textbf{Acceptable} tiers (scores between $0.6$ and $0.9$). This high-confidence data was organized into three subsets:\textit{\textbf{US-Std(Standard pakistani urdu)}}, \textbf{\textit{US-EngPk(Pakistani accented english)}}, and \textbf{\textit{US-CS(code-switched).}}The data was categorized by duration: \textit{Short} ($\le$ 10s) and \textit{Long} (10--35s).
 
\section{Human-centric quality assessment}

\subsection{Experimental setup and recruitment}
To validate the corpus, 180 clips across three sets (A, B, and C) were randomly sampled by complexity using an anchor set strategy (Table~\ref{tab:validation_sampling}). Six university-recruited native Urdu speakers (3M/3F) evaluated the data in a controlled laboratory setting. To ensure independent, high-quality judgment, annotators worked in isolated pairs with mandatory 20-minute breaks between levels to prevent cognitive fatigue. This was conducted under written informed consent..
\begin{table}[h]
\centering
\caption{Sampling Strategy for Human Quality Assessment.}
\label{tab:validation_sampling}
\scriptsize 
\setlength{\tabcolsep}{4pt} 
\begin{tabular}{@{}l c c c@{}}
\toprule
\textbf{Complexity Stratification} & \textbf{Anchor} & \textbf{Unique} & \textbf{Total} \\ 
\textbf{(Per Set A, B, C)} & \textbf{(Common)} & \textbf{(per Set)} & \textbf{(n=180)} \\
\midrule
US-Std,US-EngPk(Short $\le$10s)& 7 & 13 & 60 \\
US-CS(Short $\le$10s)& 7 & 13 & 60 \\
Mixed (Long 10--35s)         & 6 & 14 & 60 \\
\textbf{Total per Set}       & \textbf{20} & \textbf{40} & \textbf{60} \\
\bottomrule
\end{tabular}
\end{table}

\subsection{Assessment framework}
Our evaluation utilized a 5-point Likert scale to measure seven key dimensions: audio quality, transcription accuracy, demographics (age, gender, accent), prosody, affect, articulation, and contextual accuracy. This multidimensional Mean Opinion Score (MOS) follows ITU-T P.800 protocols \cite{itu1996recommendation}. We also collected open-ended feedback to catch specific errors like misspellings or omissions.

\subsection{Evaluation of annotator responses}
Quantitative analysis was performed using Pandas \cite{mckinney2010data}, with inter-rater reliability (IRR) validated via Cohen’s $\kappa$ \cite{cohen1960coefficient} and Fleiss’ $\kappa_f$ \cite{fleiss1971measuring} through scikit-learn \cite{pedregosa2011scikit} and statsmodels \cite{seabold2010statsmodels}. To address subjectivity, we calculated exact and adjacent ($\pm 1$) Inter-Annotator Agreement (IAA) across three unique sets and a shared anchor set.

Results confirm the corpus's high fidelity (Mean MOS: 4.64, $\sigma = 0.74$), with 92.78\% of ratings being 4s or 5s. While Cohen’s $\kappa$ reached 0.678 (Set B) and 0.545 (Set C), the global $\kappa_f$ of 0.141 illustrates the "Kappa Paradox." The lack of variance in a consistently high-quality dataset suppresses $\kappa_f$ despite a robust 87.67\% adjacent IAA. As shown in Figure~\ref{fig:assessment_results} A, one strict annotator (Mean 4.12) vs. others (up to 4.95) reflects natural perceptual diversity; this variance lowers global Kappa while maintaining high overall consensus. 

\begin{figure}[htbp]
  \centering
  \includegraphics[width=\linewidth]{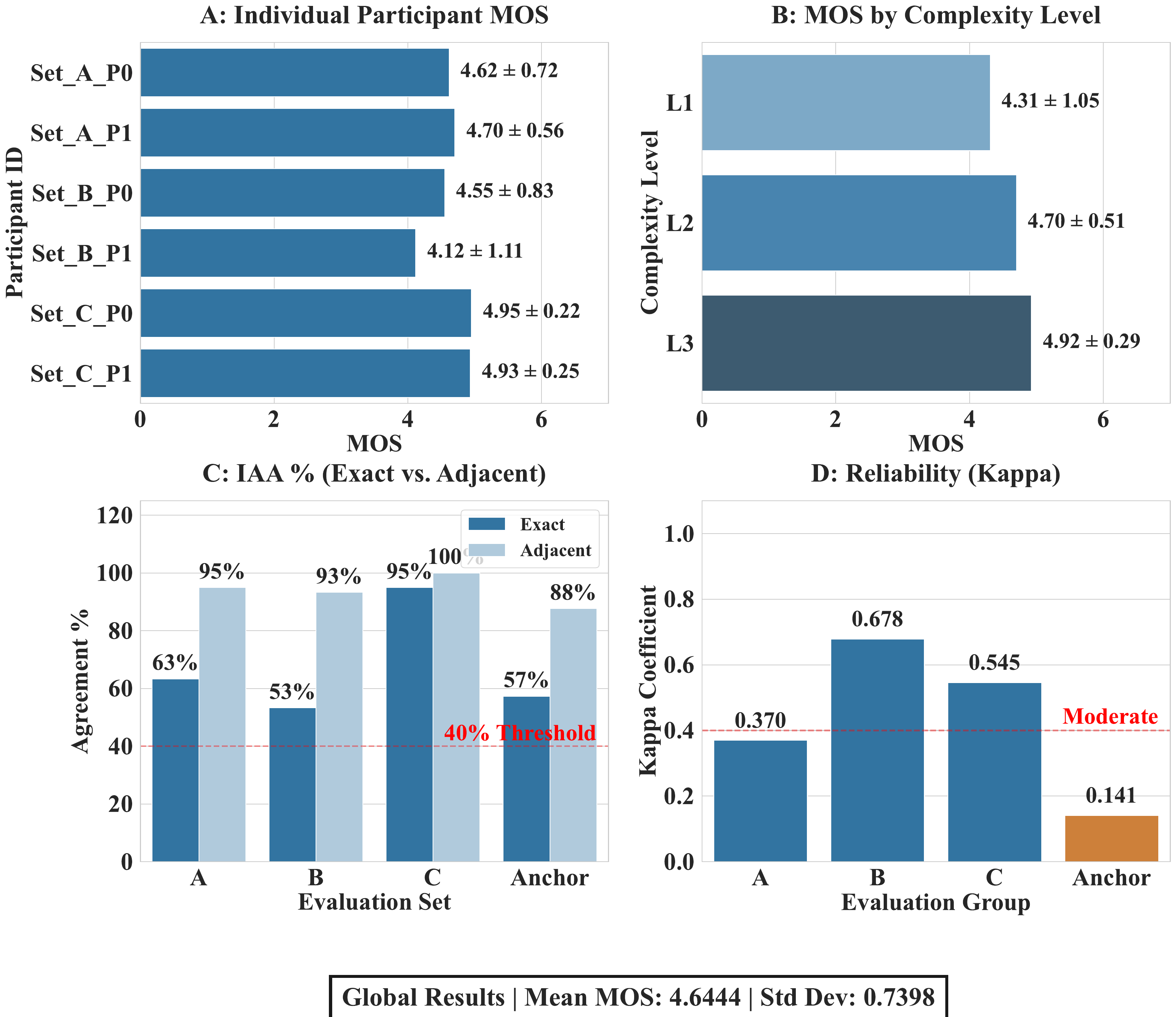}
  
  \caption{Detailed results for human-centric assessment}
  \label{fig:assessment_results}
\end{figure}

\section{UrduSpeech corpus}
\begin{table*}[t]
\caption{Comparison of Our corpus with Existing Urdu and Multilingual Datasets.}
\label{tab:dataset_comparison}
\centering
\small
\setlength{\tabcolsep}{1pt}
\begin{tabular}{l c c c l l l}
\toprule
\textbf{Dataset} & \textbf{Year} & \textbf{Hours} & \textbf{Speakers} & \textbf{Linguistic Focus} & \textbf{Metadata} & \textbf{Access} \\
\midrule
ARL Urdu (LDC) \cite{LDC_ARL_Urdu_2007} & 2007 & 20 & 200&  Read& None & Paid (\$4k) \\
CLE Spontaneous \cite{cle_urdu_speech_2016}& 2016& 0.86& 40& Spontaneous Urdu & single word utterance & Licensed\\
Common Voice (Mozilla)\cite{ardila2020common} & 2020 & 81*& 498& General ASR & Age/Gender & Open \\
Google FLEURS\cite{conneau2023fleurs} & 2023 & 12 & 100*& Benchmarking & Transcription & Open \\
LDC-IL Urdu\cite{LDCIL_Urdu_2023} & 2023 & 50 & 434& Regional Dialects & Transcription & Licensed \\
Urdu-Bench \cite{arif2025we} & 2025 & 1.3& 10& Conversational & Benchmarking & Licensed \\
\midrule
\textbf{UrduSpeech(Ours)}& \textbf{2026} & \textbf{156} & \textbf{1,000+*}& \textbf{Urdu-Std,Eng-Pk,CS} & \textbf{Transcription,12 Para.Labels} & \textbf{Free / Open} \\
\bottomrule
\end{tabular}
\end{table*}
\subsection{Corpus Distribution and Statistics}
The UrduSpeech corpus comprises \textbf{91GB} 156 hours of diarized audio. As shown in Figure~\ref{fig:Category_distribution}, the Interview category represents the largest share, accounting for approximately 34 hours (21\% of the total volume). Traditional genres such as drama and poetry contain a higher volume of Us-Std, whereas conversational categories including interviews, podcasts, and vlogs feature a majority of US-CS data.

\begin{figure}[htbp]
\centering
\includegraphics[width=\linewidth]{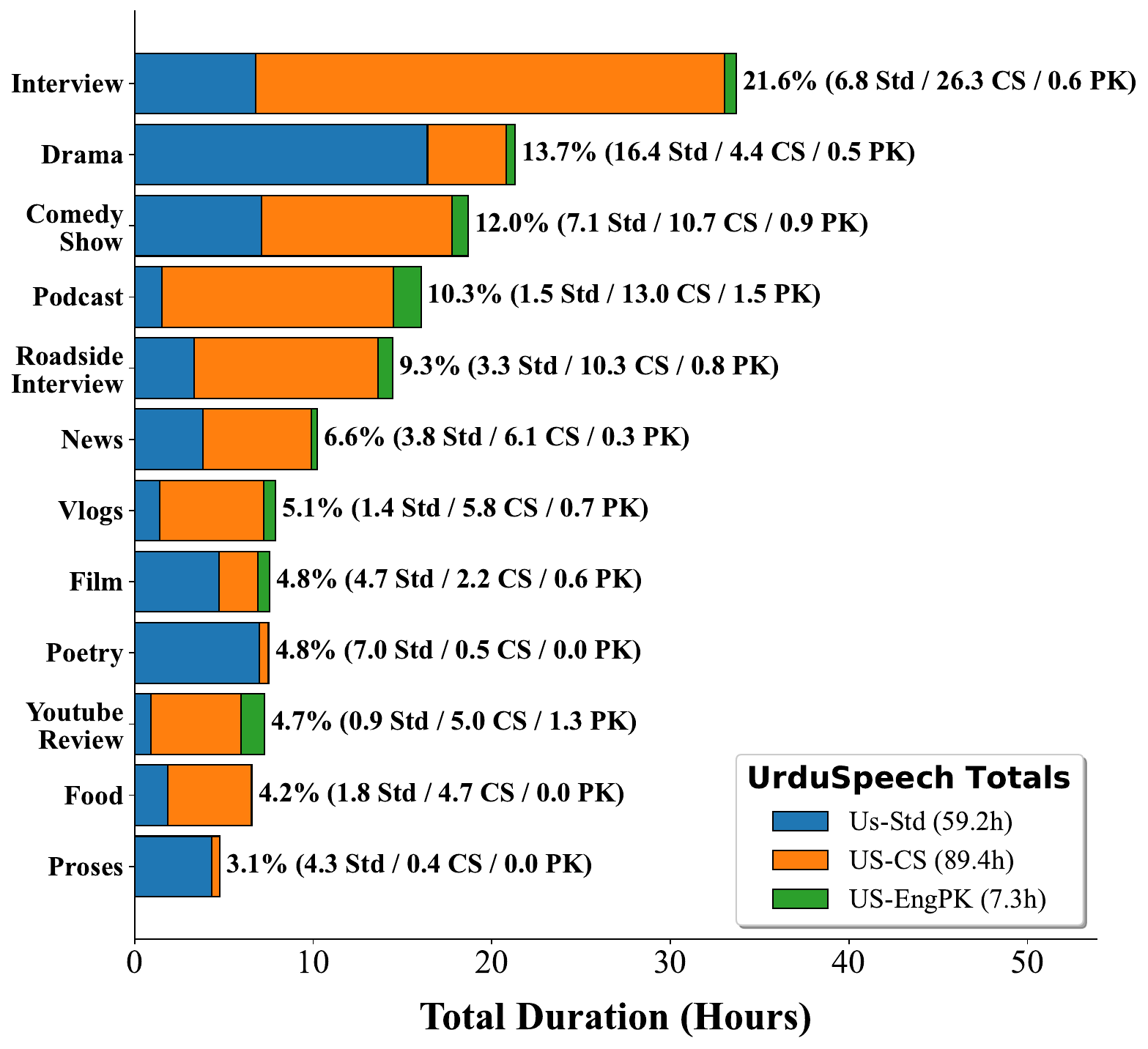}
\caption{Corpus data distribution across subsets and categories}
\label{fig:Category_distribution}
\end{figure}

The corpus contains 71,792 diarized segments, categorized by duration into short-format (55,407 segments) and long-format (16,243 segments) clips. Detailed demographic and linguistic insights are provided in Table~\ref{tab:corpus_stats}. While age demographics trend toward young adults and middle-aged speakers, there is a notable presence of children and the elderly.Notably, the average duration, word count, and words per second (WPS) for US-CS data are higher than for US-Std. Although US-Std utterances outnumber code-switched segments by approximately 9,000, code-switched audio accounts for 30 hours more in total duration, reflecting the expansive nature of conversational speech. The data spans a diverse linguistic spectrum, including:

\begin{itemize}
\item \textbf{Conversational:} Podcasts and formal interviews.
\item \textbf{Narrative:} Prose, poetry, and monologues.
\item \textbf{Archival:} Historical dramas, essays, broadcasts, and poems.
\item \textbf{Daily Life:} Vlogs, reviews, culinary content, and roadside interviews.
\item \textbf{Informative/Entertainment:} News, comedy shows, dramas, and films.
\end{itemize}
\begin{table}[t]
\caption{Demographic and Linguistic Statistics of the Corpus.}
\label{tab:corpus_stats}
\centering
\small
\setlength{\tabcolsep}{3pt} 
\footnotesize 
\begin{tabular}{@{}llrlr@{}}
\toprule
\textbf{Category} & \textbf{Subset} & \textbf{Count} & \textbf{Metric} & \textbf{Value} \\ \midrule
Utterance    & Female        & 28,802 & Avg. Dur (Urdu) & 5.60s  \\
          & Male          & 42,990 & Avg. Dur (Eng)  & 6.05s  \\ \cmidrule(lr){1-3}
Age Group & Young Adult   & 34,126 & Avg. Dur (CS)   & 10.96s \\ \cmidrule(lr){4-5}
          & Middle Age    & 33,495 & Avg. WPS (Urdu) & 2.90   \\
          & Child         & 1,804  & Avg. WPS (Eng)  & 2.70   \\ 
          & Elderly       & 2,367  & Avg. WPS (CS)   & 3.33   \\ \cmidrule(lr){1-3} \cmidrule(lr){4-5}
Accent    & Std. Urdu     & 38,036 & Avg.Word count(Urdu) & 16.22  \\
          & Std. English  & 4,372  & Avg.Word count(Eng)  & 16.33  \\ 
          & Urdu-Eng CS   & 29,384 & Avg.Word count(CS)   & 36.50  \\ \midrule
\textbf{Total} & \textbf{Clips} & \textbf{71,792} & \textbf{Total Hours} & \textbf{156.0h} \\ \bottomrule
\end{tabular}
\end{table}

\subsection{Comparison with Existing Resources}
As detailed in Table \ref{tab:dataset_comparison},  UrduSpeech represents a significant advancement in the landscape of South Asian speech resources. While foundational datasets such as ARL Urdu \cite{LDC_ARL_Urdu_2007} and LDC-IL \cite{LDCIL_Urdu_2023} provided early benchmarks, they are often constrained by restrictive licensing, high costs, or a limited number of speakers. In contrast,Our proposed corpus offers 156 hours of high-quality audio, nearly 40\% more volume of the common-voice validate set \cite{ardila2020common}.

A critical differentiator of our corpus is its unprecedented speaker diversity. While "massively multilingual" initiatives like Google FLEURS \cite{conneau2023fleurs} and Mozilla Common Voice \cite{ardila2020common} include Urdu, they suffer from a severe scarcity of unique voices for the language, often featuring fewer than 20 validated speakers. Our corpus addresses this gap by providing data from approximately 3,000 unique speakers, ensuring robust model generalization across diverse demographics. Moreover, unlike existing datasets that provide only basic transcriptions, Our corpus is the first to integrate a 12-dimension paralinguistic metadata framework enabling multifaceted research into affective computing and speaker profiling in the context of Urdu-English code-switching.
\section{Limitation and future work}
Our corpus provides a substantial resource for Urdu, code-switched Urdu-English, and Pakistani-accent English speech research, yet several limitations exist. First, while automated diarization via Pyannote 3.1 identified over 3,000 unique speaker clusters, we conservatively estimate the count at 1,000+ unique speakers to account for a potential machine error margin of approximately 2,000 clusters due to over-segmentation in "in-the-wild" recordings. While the gender distribution across utterances has been manually verified, ongoing work is dedicated to validating unique speaker IDs to ensure absolute compliance.

Additionally, despite robust source separation via Demucs and Spleeter, some segments retain secondary speakers or background environmental noise. Future work will focus on establishing baseline benchmarks for ASR and TTS. We are currently developing a custom tokenizer and implementing forced-alignment for word-level temporal precision to enhance the corpus's utility for complex prosodic and acoustic modeling.

\section{Conclusion}
In this study, we introduced UrduSpeech, a 156-hour (91 GB) multi-domain speech corpus featuring 12-dimensions paralinguistic metadata. By developing a robust and reproducible pipeline, we successfully addressed the complexities of "in-the-wild" Urdu speech and the high prevalence of Urdu-English code-switching. Our stratified methodology resulted in a high-diversity dataset with three specialized subsets: US-Std for standard Pakistani Urdu, US-CS for code-switching research, and UA-EngPk for Pakistani-accented English.

To ensure data integrity, we implemented a rigorous human-centric validation framework. Assessment by native speakers yielded a global Mean Opinion Score (MOS) of 4.64 ($\sigma = 0.74$), with inter-annotator agreement metrics including a Cohen’s $\kappa$ exceeding 0.4 and an average exact agreement of 57\%,validating the reliability of our labels. These results, coupled with a high transcription confidence of 97.6\%, demonstrate that UrduSpeech provides high-fidelity, human-verified ground truth. We believe that this corpus, alongside our open-source pipeline, will serve as a catalyst for future research in Urdu and other under-resourced Perso-Arabic script languages.
\section{Generative AI Use Disclosure}
The authors acknowledge the use of generative AI tools solely for text refinement, grammar corrections, and proofreading of the manuscript. All technical methodologies, data collection, and original research contributions were conceived and executed entirely by the authors.
\newpage
\bibliographystyle{IEEEtran}
\bibliography{mybib}
\end{document}